\documentclass[useAMS,usenatbib]{mn2e}

\usepackage{times}
\usepackage{graphics,epsfig}
\usepackage{graphicx}
\usepackage{amsmath}
\usepackage{amssymb}
\usepackage{bm}
\usepackage{dcolumn}
\usepackage{epsfig}
\usepackage{subfig}
\usepackage{tabularx}
\usepackage[usenames,dvipsnames,svgnames,table]{xcolor}

\newcommand{\kms}{km~s$^{-1}$}
\newcommand{\acc}{\ensuremath{\rm cm^{-2}}}
\newcommand{\ms}{\ensuremath{\rm M_{\odot}}}

\newcommand{\msyr}{\ensuremath{\rm M_{\odot} \rm yr^{-1}}}
\newcommand{\arcs}{\ensuremath{^{\prime\prime}}}
\newcommand{\co}[1]{{\color{Cerulean}{#1}}}

\title[H{\sc i} mapping of the host of GRB 980425]
{First measurement of H{\sc i}~21\,cm emission from a GRB host galaxy indicates a post-merger system}
\author[Arabsalmani et al.]
{Maryam Arabsalmani$^{1,2}$\thanks{E-mail: marabsal@eso.org}, Sambit Roychowdhury$^3$,  Martin Zwaan$^1$, Nissim Kanekar$^4$,
\newauthor
Micha\l ~J. Micha\l owski$^5$
       \\
        $^1$European Southern Observatory, Karl-Schwarzschild-Strasse 2, 85748 Garching bei M\"{u}nchen, Germany\\
	$^2$Dark Cosmology Centre, Niels Bohr Institute, University of Copenhagen, Juliane Maries Vej 30, DK-2100 Copenhagen \O, Denmark\\
	$^3$Max-Planck-Institut f\"{u}r Astrophysik, Karl-Schwarzschild-Strasse 1, 85748 Garching bei M\"{u}nchen, Germany\\
        $^4$National Centre for Radio Astrophysics, Tata Institute of Fundamental Research, Post Bag 3, Ganeshkhind, Pune 411 007, India\\
        $^5$Institute for Astronomy, University of Edinburgh, Royal Observatory, Blackford Hill, Edinburgh EH9 3HJ, UK
}

\begin{document}
\date{}

\pagerange{\pageref{firstpage}--\pageref{lastpage}} \pubyear{}

\maketitle

\label{firstpage}

\begin{abstract}

We report the detection and mapping of atomic hydrogen in H{\sc i}~21\,cm emission from ESO 184-G82, the host 
galaxy of the gamma ray burst 980425. This is the first instance where H{\sc i} in emission has been detected from a 
galaxy hosting a gamma ray burst. ESO 184-G82 is an isolated galaxy and  contains a Wolf-Rayet 
region close to the location 
of the gamma ray burst  and the associated supernova, SN 1998bw.  This is 
one of the most luminous H{\sc ii} regions identified in the local Universe, with a  
very high inferred density of star formation. 
The H{\sc i}~21\,cm observations reveal a high H{\sc i} mass for the galaxy, twice as large as the stellar mass.
The spatial and velocity distribution of the H{\sc i}~21\,cm emission reveals a disturbed rotating gas disk, which suggests 
that the galaxy has undergone a recent minor merger that disrupted its rotation.
%
%
We find that the Wolf-Rayet region and the gamma ray burst are both located in the highest H{\sc i} column density region of the galaxy. We speculate that the merger event has resulted in shock compression of the gas, triggering extreme star formation activity, and resulting in the formation of both the Wolf-Rayet region and the gamma ray burst. 
The high  H{\sc i} column density  environment of the GRB is consistent 
with the high  H{\sc i} column densities seen in absorption in the host galaxies of high redshift gamma ray bursts.

\end{abstract}


\section{Introduction}

Long-duration gamma ray bursts (GRBs) are believed  to originate in the death of short-lived massive 
stars and hence are expected to be located in regions with high star formation \citep{Bloom02, Fruchter06}. This picture 
is in good agreement with the high star formation rates (SFRs) typically observed in galaxies hosting GRBs. 
While determining the stellar mass and the SFR of GRB host 
galaxies at high redshifts remains challenging even with today's 10m-class telescopes, 
such estimates have been possible for a fair number of GRB hosts out to $z \sim 3$ 
\citep[e.g.][]{
Castroceron10} 
This remarkable recent progress in studies of GRB host galaxies has not been 
mirrored in emission studies of their neutral gas, which is the fuel for star formation. 
Such studies are critical in order to understand the interplay between interstellar 
medium (ISM) conditions and star formation that gives rise to the GRB progenitors. 
The neutral atomic hydrogen in several GRB host galaxies at ${\rm z}\gtrsim \rm 2$ has been 
detected in 
absorption, yielding estimates of 
the H{\sc i} column density, N(H{\sc i}),   
and gas kinematics 
\citep[e.g.][]{Fynbo09, Prochaska08, Arabsalmani15}. 
Indeed, molecular hydrogen has also been detected in absorption in three GRB 
hosts \citep{Prochaska09, Kruhler13, Delia14}. 
However, these absorption features only trace the gas along the narrow beam 
in the GRB sightline and carry little information on the whole galaxy. 
Understanding the nature of GRB host galaxies and the conditions for GRB formation 
critically requires emission studies in the atomic and molecular gas. 
Although CO emission has recently been detected in three galaxies hosting GRBs 
\citep{Hatsukade14, Stanway15}, 
uncertainties in the CO-to-H$_2$ conversion factor \citep[e.g.][]{Bolatto13} imply large 
uncertainties in the inferred molecular gas mass. 
Unfortunately, 
the sensitivity of today's radio telescopes limits H{\sc i}~21\,cm emission studies 
to relatively low redshifts, $z \lesssim 0.2$ \citep[e.g.][]{Catinella15}. 
To date, the information on the H{\sc i} mass, which is likely to be the dominant part of the gas content 
of such galaxies, does not exist for any GRB host. The H{\sc i}~21\,cm emission line would allow a direct measurement of  
the total H{\sc i} mass, as well as detailed studies of the spatial distribution and the kinematics of the atomic gas. 
Combining this information on the H{\sc i} content, distribution and kinematics, with information 
on the star formation and the stellar mass, and, finally, with information on the molecular 
gas,  
would enable us to have a comprehensive picture of galaxies hosting GRBs. 

The galaxy ESO 184-G82, the host of the closest known GRB  \citep[at $z=0.0087$;][]{Foley06} and one of the first GRBs to be associated with a supernova \citep[SN 1998bw,][]{Galama98}, 
is one of the few GRB hosts where it is possible to carry out spatially-resolved spectroscopy 
and photometry. It hence offers the unique opportunity of a detailed study of the close environment of a GRB. 
The GRB occurred in one of the several high surface brightness star-forming 
regions of the galaxy \citep{Fynbo00}. 
The host galaxy has a high specific star formation rate (SFR per unit stellar mass, sSFR); 
in particular, it contains a Wolf-Rayet (W-R) region  with extremely high ongoing star formation, close to the GRB location.
The cause of this high star formation 
has been an unsolved puzzle over the last decade.  
In this Letter, we present evidence for the possible cause of the extreme star formation properties of ESO\,184-G82, via a study of the H{\sc i} 21\,cm line emission from this galaxy.
This is the first time that H{\sc i} has been detected in emission from a GRB host galaxy (see also Michalowski et al., in press). 
\begin{figure}
\begin{center}
\psfig{file=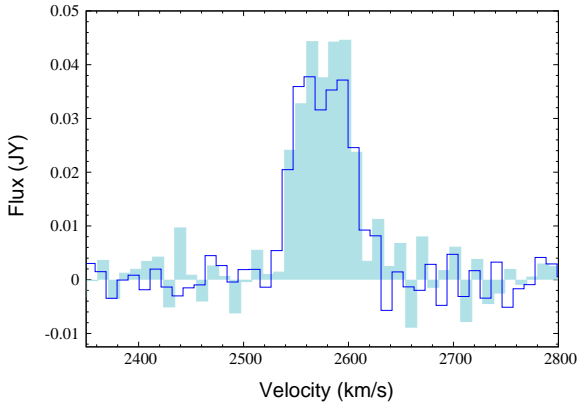,width=3.1truein}
\end{center}
\vskip -3 mm
\caption{H{\sc i} 21cm emission profile of the galaxy ESO~184-G82, as measured with the GMRT (shaded histogram) and the ATCA (bold line). 
}
\label{fig:spec}
\end{figure}

\section {Observations, data analysis and results}
We initially obtained archival Australia Telescope Compact Array (ATCA) data  
covering the redshifted H{\sc i} 21\,cm line from the GRB\,980425 host 
galaxy (project code: C2700). These observations were carried out on 2012 April  
12 in the 1.5B configuration, with a total on-source time of 8 hours,  
using the Compact Array Broadband Backend with a bandwidth of 3.5 MHz.   
The ATCA data were  
analysed using ``classic'' {\sc aips}, following standard procedures, and 
yielded a detection of H{\sc i} 21\,cm emission from the GRB host galaxy. 
Since the ATCA has only six antennas, and hence relatively poor U-V coverage in 
a single configuration, we followed up this detection with mapping observations 
with the Giant Metrewave Radio Telescope (GMRT). 
The GMRT H{\sc i} 21\,cm observations of the GRB host galaxy were carried out 
on 2015 April 3 and 5, using a bandwidth of 4 MHz, centred at the redshifted 
H{\sc i} 21\,cm line frequency of 1410.33~MHz and sub-divided into 512~channels. 
The total on-source time was 2.6 hours, with observations of the 
standard calibrators 3C48 and 2005-489 used to calibrate the flux density  
scale, the system bandpass and the system gain. The initial data editing and  
calibration of the GMRT data were carried out in the {\sc flagcal} software 
package \citep{Prasad12}, with the remaining analysis done in 
{\sc aips}, again following standard procedures. After Hanning smoothing, re-sampling, 
and a detailed self-calibration 
process, the radio continuum was subtracted from the calibrated visibilities 
using the tasks {\sc uvsub} and {\sc uvlin}, and the residual visibilities 
mapped with different U-V tapers to produce spectral cubes at different 
spatial and velocity resolutions. The velocity resolution was varied in 
order to improve the statistical significance of the detected H{\sc i} 21\,cm 
emission in independent velocity channels. The properties of the spectral 
cubes that will be used here are listed in Table~\ref{tab:obs}. 
\begin{figure}
\begin{center}
\hskip -1 cm
\begin{tabular}{c}
{\mbox{\includegraphics[width=3.0truein]{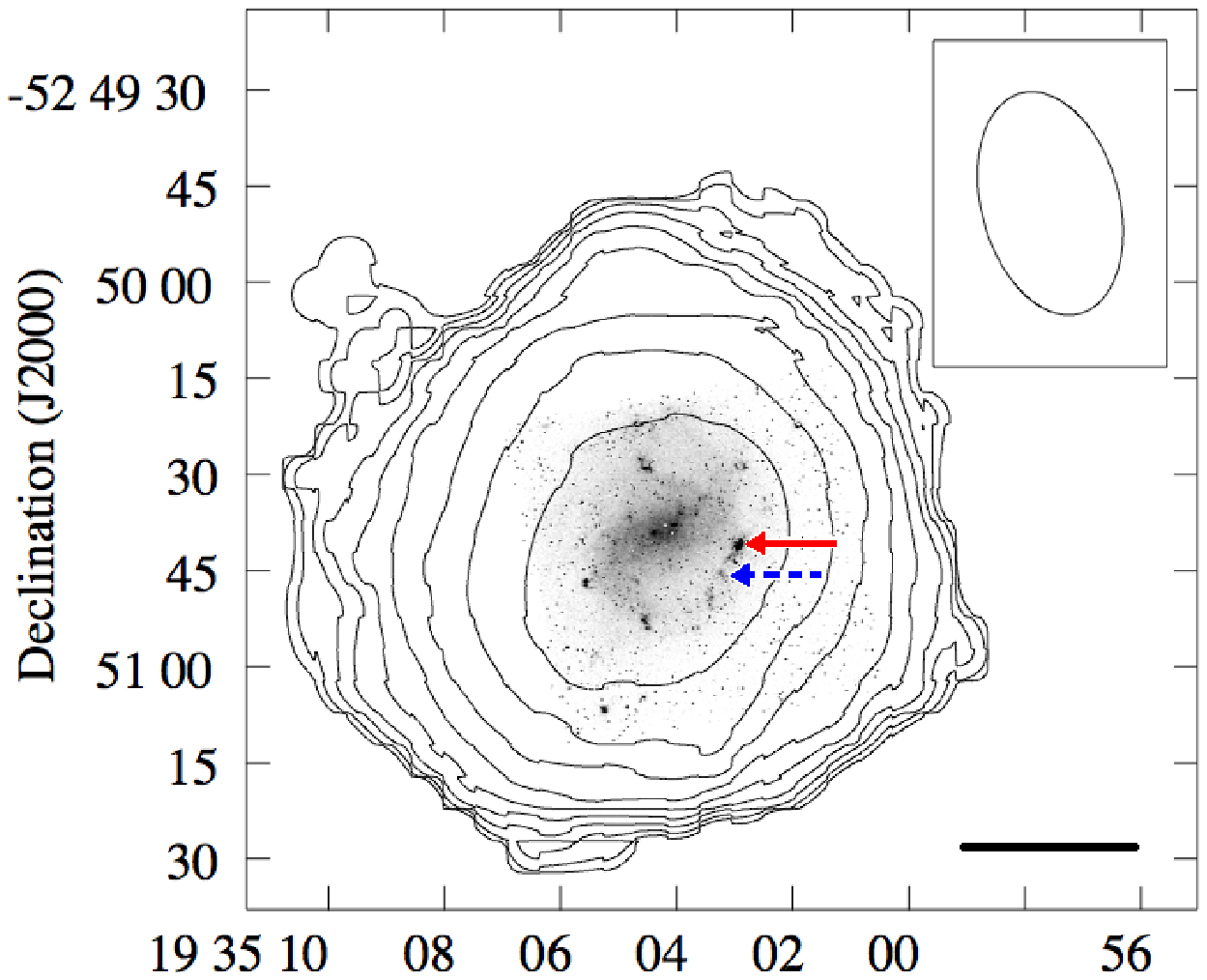}}}\\
{\mbox{\includegraphics[width=3.0truein]{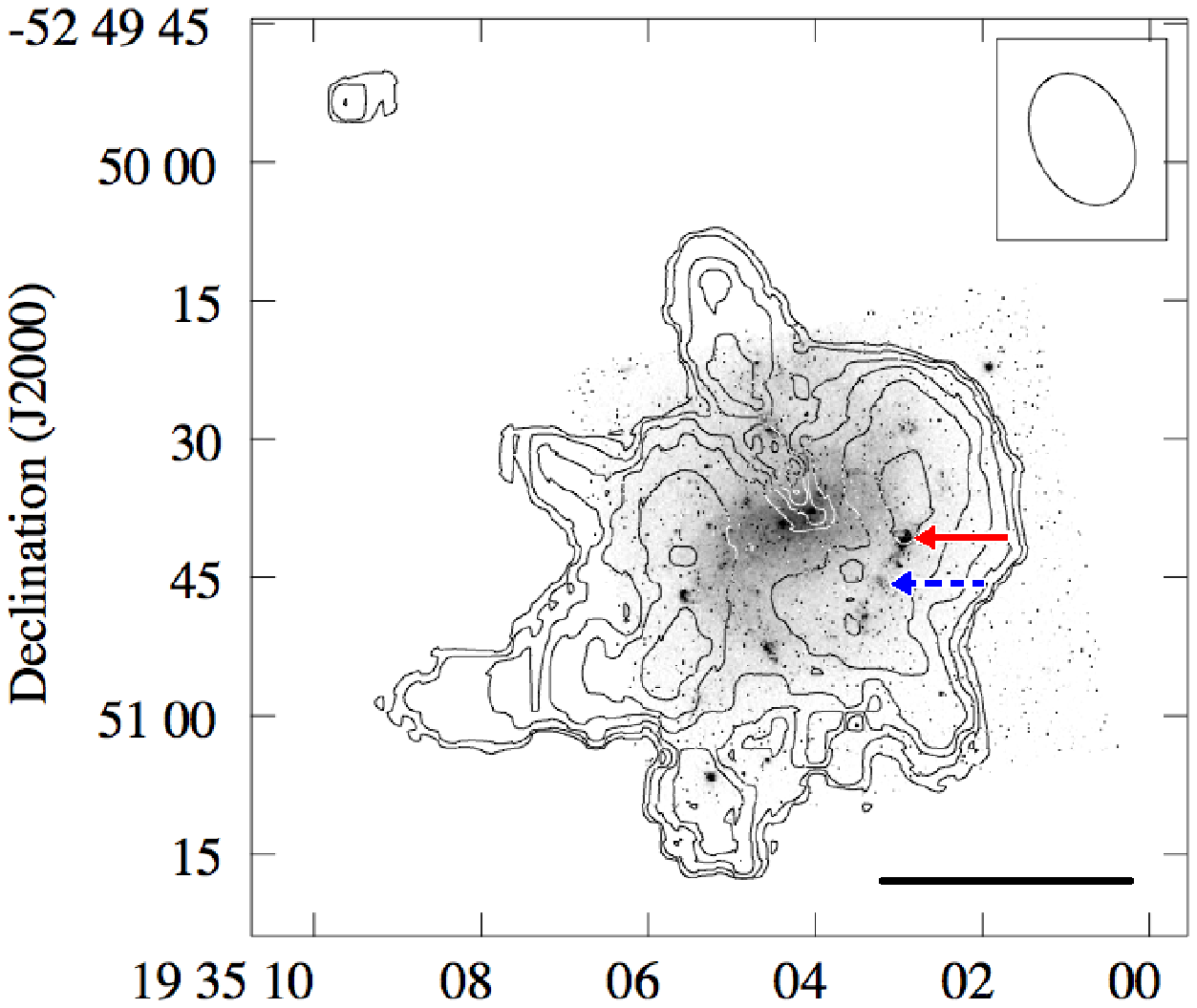}}}\\
{\mbox{\includegraphics[width=3.0truein]{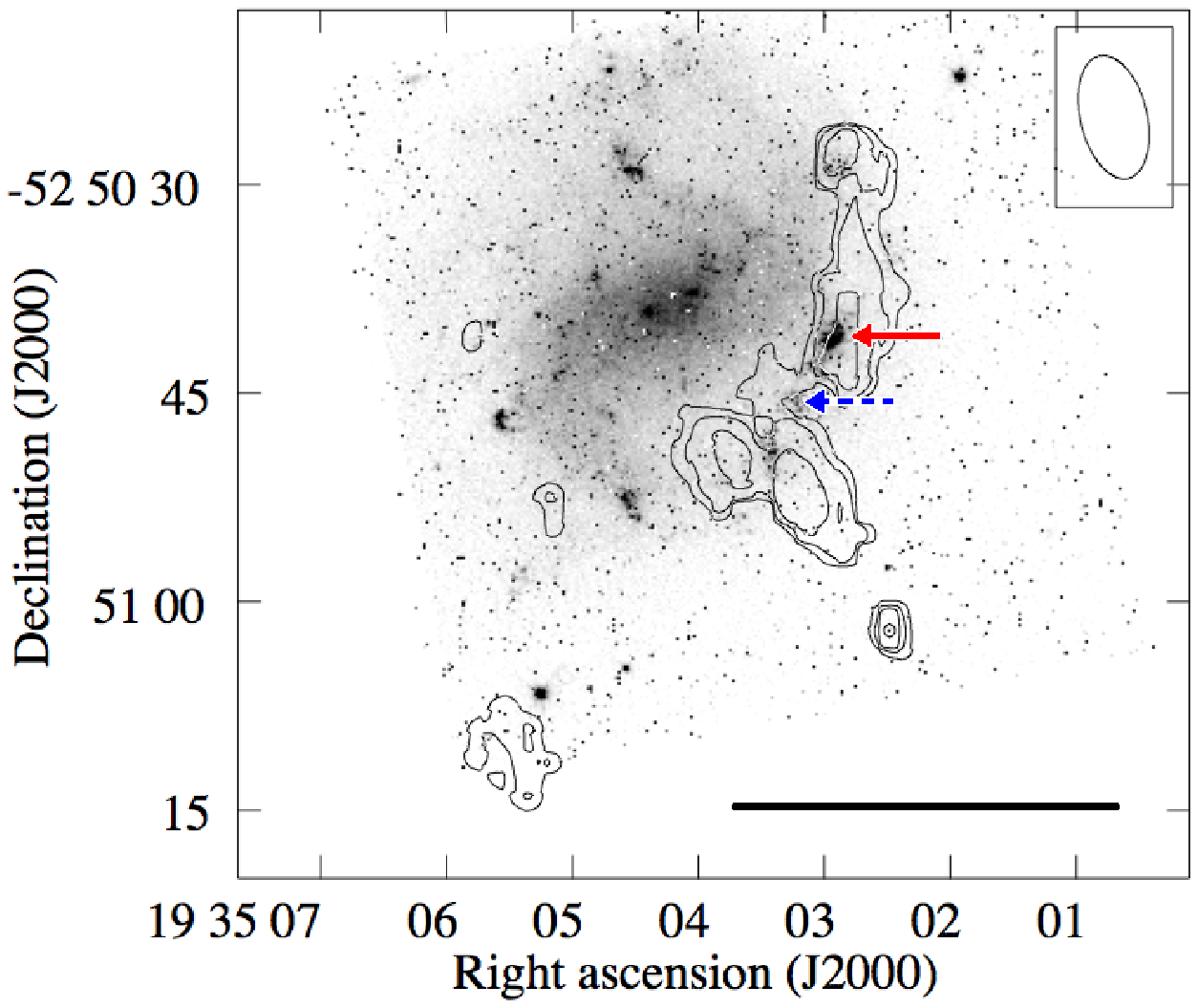}}}\\
\end{tabular}
\end{center}
\caption{
H{\sc i} column density maps (in contours) overlaid on an HST image (greyscale) of ESO~184-G82. The resolutions are 
36\arcs$\times$ 22\arcs\ (top), 15\arcs$\times$ 11\arcs\ (middle), and 9\arcs$\times$ 5\arcs\ (bottom), with outermost (3 $\sigma$) contours of 5$ \times {\rm 10^{19}}$~cm$^{\rm -2}$, 2$ \times {\rm 10^{20}}$~cm$^{\rm -2}$, and 5$ \times {\rm 10^{20}}$~cm$^{\rm -2}$, respectively, and subsequent contours at N(H{\sc i}) intervals of $\sqrt{\rm 2}$. For reference, the solid line at the bottom right of each panel indicates a scale of 5 kpc at the distance of the galaxy. The dashed blue and solid red arrows mark the GRB location and the W-R region, respectively.}
\label{fig:maps}
\end{figure}

Fig.~\ref{fig:spec} shows the H{\sc i} 21\,cm emission spectra obtained from the ATCA  
and GMRT spectral cubes, at, respectively, spatial resolutions of 34.5\arcs$\times$ 21.4\arcs\ and  
36\arcs$\times$ 22\arcs\, and velocity resolutions of 10.5 \kms\ for both datasets. Note that the spectra show the integrated 
flux density over the entire spatial extent of the detected H{\sc i} 21\,cm 
emission. We have checked that all of the source flux density is recovered 
at this resolution; further lowering the resolution does not increase the 
total flux density. The ATCA and GMRT spectra are seen to be in 
agreement; 
since the GMRT data have far better U-V coverage than the ATCA data, and also yield a significantly better angular resolution, all the following results and discussion will be based on the GMRT data. 
We obtain an integrated H{\sc i} 21\,cm line flux density of 3.0 $\pm$ 0.2 Jy \kms ~from the GMRT spectrum of Fig.~\ref{fig:spec}. 
Using a source distance of 37.7~Mpc (assuming a flat $\Lambda$-cold dark matter cosmology with $\rm{H}_{\rm 0}$~=~69.6 km~s$^{-1}$Mpc$^{-1}$ and $\Omega_{\rm {m}}$~=~0.3), this yields 
an H{\sc i} mass of (1.00 $\pm$ 0.08) $\times$~10$^{\rm 9}$~\ms.
Using the coarsest resolution H{\sc i}~21\,cm map we measure the inclination angle of the H{\sc i} disk to be between $42^\circ$ and $50^\circ$ for an intrinsic 
axial ratio varying between 0.1 and 0.5. This range in inclination is consistent  with the inclination angle of the optical disk, $i = 50^\circ$ \citep{Christensen08}. 
The velocity width between half-power points is W$_{\rm 50}$~$=$~65 \kms.  Correcting this for the 
inclination angle of $i = 50^\circ$ yields a velocity width of W$_{\rm 50}^{\rm i}$~$=$~ 85 \kms. 
The {\sc aips} task {\sc momnt} was used to make H{\sc i} column density maps of the field, using the 7 \kms\ resolution data cubes, at different spatial resolutions. Fig.~\ref{fig:maps} shows three of these maps (in contours) overlaid on an Hubble Space Telescope (HST) optical image (with the {\small MIRVIS} filter, centred at 5737.453$\AA$; in greyscale). The top panel shows that the H{\sc i} disk of the galaxy is more extended than the optical disk, a feature common in sub-L$_*$, gas-rich galaxies \citep[e.g.][]{Begum08}.
The broad peak of the H{\sc i} distribution at the coarsest resolutions corresponds to the location of the optical galaxy.
The intermediate-resolution (15\arcs$\times$ 11\arcs) map, in the middle panel of Fig.~\ref{fig:maps}, shows an arc of dense gas roughly coincident with the southern spiral arm of the galaxy. The W-R region mentioned above arises in one of the denser H{\sc i} regions. There appears to be a lack of high-N(H{\sc i}) gas at the centre of the galaxy, as well as towards the north-east of the optical disk. An apparently isolated high-N(H{\sc i}) knot is visible at the extreme north-east end of the H{\sc i} disk; its signature 
is clear in even the low resolution map. 
The highest resolution (9\arcs$\times$ 5\arcs) map, shown in the bottom panel of Fig.~\ref{fig:maps}, is sensitive to only the highest H{\sc i} column densities. This image confirms that both the W-R region and the GRB location are coincident with the highest N(H{\sc i}) regions of the galaxy. Indeed, when inspecting the N(H{\sc i}) maps at higher resolution, we notice a piling up of gas on the western side of the galaxy. 

\begin{figure}
\begin{center}
\hskip -1 cm
\psfig{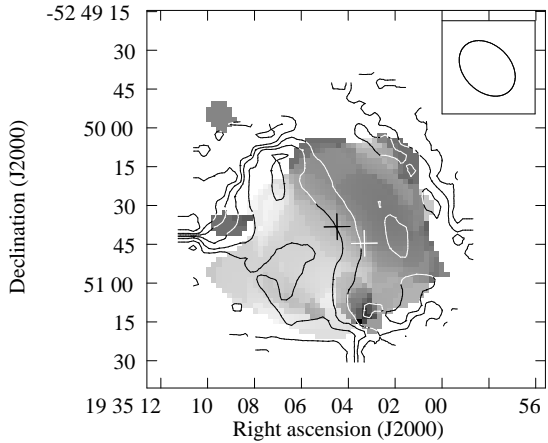}
\end{center}
\caption{
The H{\sc i}~21\,cm velocity field of ESO~184-G82, at two resolutions, 36\arcs$\times$ 22\arcs (contours, ranging, in intervals of 7  \kms, from 2553 \kms\ at the bottom left to 2609 \kms\ at the top right) and 25\arcs$\times$ 18\arcs\ (greyscale, from 2550 \kms\ as lightest  to 2630 \kms\ as darkest). Crosses mark the locations of the galaxy's optical centre and the GRB.}
\label{fig:vel}
\end{figure}

To fully understand the nature of the H{\sc i} distribution in the galaxy, one needs to study the velocity distribution of the gas.
The intensity-weighted H{\sc i}~21\,cm velocity field of the full H{\sc i} disk, displayed in contours in Fig.~\ref{fig:vel} at a low resolution (36\arcs$\times$ 22\arcs), shows the presence of an overall gradient from the south-east to the north-west. 
However, even at this coarse resolution, there are multiple features incompatible with an origin in a regularly rotating disk galaxy. For a more detailed study of the velocity field, we use the data cube of resolution 25\arcs$\times$ 18\arcs, which allows us to spatially distinguish different H{\sc i} emission regions, while retaining some sensitivity to low N(H{\sc i}) gas. The H{\sc i}~21\,cm velocity field at this resolution is shown in greyscale in Fig.~\ref{fig:vel}. The rotation of the neutral gas around the galaxy's optical centre is evident at this resolution; the velocity field traced by H-$\alpha$ emission in a limited region around the optical centre \citep{Christensen08} is consistent with it.
\begin{table}
\caption{Parameters of the GMRT H{\sc i} data cubes}
\label{tab:obs}
\begin{tabular}{cccc}
\hline
Synthesized Beam&Channel width&Noise in line-free channel\\
(arcs$\times$arcs)&(\kms)&(mJy Bm$^{\rm -1}$)\\
\hline
35.6$\times$21.5&10.5~,~7.0&1.5~,~1.8\\
24.6$\times$18.1&7.0&1.6\\
15.0$\times$10.5&7.0&1.3\\
~9.1$\times$~4.8&7.0&1.0\\
\hline
\end{tabular}
\end{table}

Fig.~\ref{fig:chans} shows the H{\sc i}~21\,cm emission from individual (7 \kms) velocity channels of the 
25\arcs$\times$ 18\arcs\ data cube. A number of H{\sc i}~21\,cm emission regions with velocities unrelated to the main rotational gradient of the H{\sc i} distribution are easily identified in the channel maps. Table~\ref{tab:knots} lists the locations of four such regions, conservatively identified in at least two velocity channels. The previously mentioned isolated knot of 
high-N(H{\sc i}) gas is also identified and marked (region c).  
For each region, we have identified the velocity range over which the ``offset'' H{\sc i}~21\,cm emission is detected (listed in column 3 of Table~\ref{tab:knots}). We use the fluxes from the ``offset'' regions in velocity channels where they are detected to estimate their H{\sc i} masses. Column 4 of the table lists the mass fraction of each region (relative to the total H{\sc i} mass of ESO~184-G2, measured at the same spatial resolution). These estimates provide a conservative lower limit to the total amount of disturbed gas in the galaxy, as they only include the mass of regions that are spatially distinct from the gas undergoing regular rotation and also only include regions identified in more than one velocity channel. Summing the mass fractions of the four kinematically disturbed regions, we find that at least 21\% of the H{\sc i} in ESO~184-G2 appears to not be following the rotation of the main gas disk. The largest H{\sc i} region with disturbed kinematics has $\sim$12\% of the total H{\sc i} mass of the galaxy, and is located close to the south-eastern peak of the main body of the rotating gas. To clearly discern the presence of the large amount of disturbed H{\sc i} close to the rotating disk, we show a position-velocity cut along the major axis of the galaxy in  Fig.~\ref{fig:pvcut}. This  kinematically-disturbed H{\sc i} is centred at an angular offset of $-$30\arcs\ from the galaxy centre and is clearly separate from the gradient representing the main rotating gas in ESO~184-G82.

\section{Discussion and conclusions}

The GRB host galaxy ESO 184-G82,  a barred spiral galaxy,  
is a low-luminosity object  with ${\rm L_B}={\rm 0.05L_B}^*$,  
but is clearly undergoing active star formation
\citep{Fynbo00, Sollerman05}. The overall SFR 
and dust properties of the galaxy are consistent with those of local dwarf galaxies 
\citep{Michalowski09}. 
The galaxy has an oxygen abundance of 0.41 solar
\citep{Sollerman05, Christensen08} 
and a stellar mass of 
4.8~$\times$~10$^{\rm 8}$~\ms\ \citep{Michalowski14}. 
The HST image of the galaxy shows that its optical appearance is dominated by 
several high surface brightness star-forming 
regions, especially in the southern spiral arm of the galaxy. 
The GRB occurred in one of these H{\sc ii} regions, 
whose properties (e.g. SFR, reddening, stellar mass) are similar to those of other H{\sc ii} regions in the galaxy
\citep{Christensen08}. 

We find  ESO 184-G82 to be a gas-rich galaxy, with an H{\sc i} mass  $\sim 2.1$ times its stellar mass, which     
is consistent with the above  studies showing ongoing star formation. 
Its SFR estimates, determined from H-$\alpha$ and U-V emission, lie in the range 0.25 - 0.45 \msyr \citep{Sollerman05, Christensen08, Michalowski09, Castroceron10}. The H{\sc i} mass and SFR of ESO~184-G82 are consistent with the relation between the two quantities in nearby, H{\sc i}-selected galaxies \citep{doy06}.
Similarly, the sum of the H{\sc i} and the stellar mass, and the H{\sc i}~21\,cm velocity width are consistent with the baryonic Tully-Fisher relation determined for nearby galaxies
\citep{zar14}. 
Thus, ESO 184-G82 appears to be a fairly typical star-forming galaxy, based on its global H{\sc i} properties.
\begin{figure}
\begin{center}
\hskip -1 cm
\psfig{file=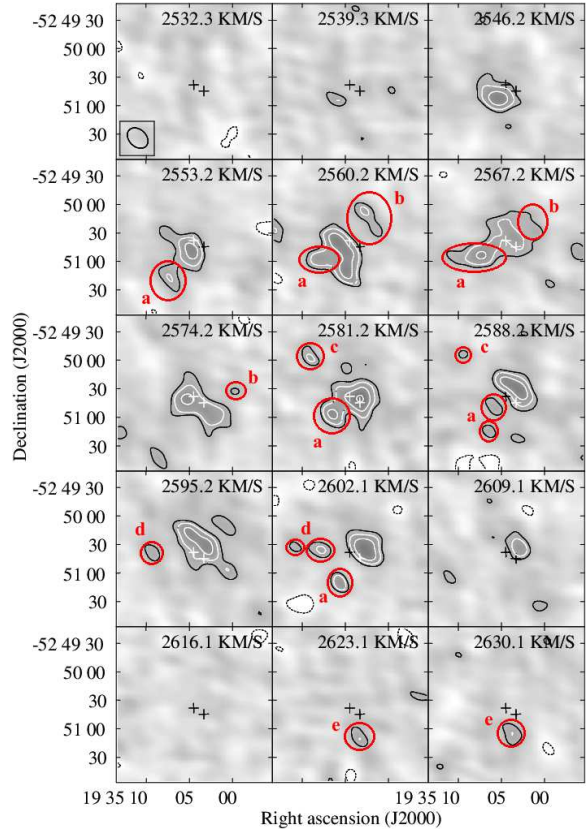,width=3.0truein}
\end{center}
\caption{
H{\sc i}~21\,cm emission from individual ( 7 \kms) velocity channels at 25\arcs$\times$ 18\arcs\ resolution. Crosses mark the locations of the galaxy's optical centre and the GRB. The outermost positive (solid) contour is at 2.5 $\sigma$ level (7$ \times {\rm 10^{19}}$~cm$^{\rm -2}$), with each subsequent contour spaced at intervals of $\sqrt{\rm 2}$; the negative (dashed) contours are at -2.5 $\sigma$ level. Regions with offset velocity (see main text) are marked in the channels in which they appear, following the nomenclature of Table~\ref{tab:knots}.
}
\label{fig:chans}
\end{figure}
\begin{table}
\caption{Offset H{\sc i} regions}
\label{tab:knots}
\begin{tabular}{ccccc}
\hline
Region&RA&Dec&Channel spread&Percentage\\
&(h m s)&(d $^\prime~^{\prime\prime}$)&(\kms)&mass\\
\hline
a&19 35 6.7&$-$52 51 04&2553.2-2602.1&12\\
b&19 35 2.1&$-$52 51 20&2560.2-2574.2&04\\
c$^ \co \dagger$&19 35 9.4&$-$52 49 55&2581.2-2588.2&02\\
d&19 35 9.0&$-$52 50 31&2595.2-2602.1&03\\
e&19 35 3.6&$-$52 51 07&2623.1-2630.1&02\\
\hline
\end{tabular}
~\\\\
$\dagger$~Region~c is spatially, not kinematically, offset from the main disk.
\end{table}

However, the unique feature of ESO~184-G82 is a region 
with a strong signature of W-R stars located at a 
projected distance of $\sim 800$ pc from the GRB location \citep{Hammer06}. 
This very young (a few Myr)  star-forming region is possibly going through its first 
episode of star formation \citep{Lefloch12},  with sSFR $\sim 11.3\rm~Gyr^{-1}$ \citep[using the SFR estimate from the H$\alpha$ emission;][]{Christensen08}, which is more 
than an order of magnitude larger than the overall  sSFR of the galaxy. 
This W-R region, many of whose properties are similar to those of high-redshift GRB host galaxies, is one of the most 
luminous and infrared-bright H{\sc ii} regions identified to date in the nearby Universe. 
It contributes substantially to the host emission at the far-
infrared, millimetre, and radio wavelengths -- something rarely observed in similar 
galaxies \citep{Lefloch06, Lefloch12, Michalowski09, Michalowski14}. 
The total infrared luminosity of the W-R region is $5 \times 10^8 \rm~ L_{\odot}$, which 
places it at the very bright end of the luminosity function of H{\sc ii} regions. Moreover, 
it has one of the highest star formation densities amongst isolated H{\sc ii} regions. Similar extra-nuclear, compact, 
star-forming complexes are rare, and usually found in starburst galaxies with far higher mass and SFR 
\citep{Lefloch12}. 
We note that it has been suggested that the progenitor of the GRB was a runaway massive star ejected from this high stellar density 
W-R region \citep{Hammer06}. 

The origin of this W-R region  has so far not been clear.
Spectroscopic observations show that the six galaxies 
in the field of ESO 184-G82 are not associated with it, and hence, interactions with them 
could not have triggered such an episode of star formation \citep{Foley06}.   
The H{\sc i}~21\,cm data, in agreement with the findings from optical spectroscopy, do not show any evidence of a large companion that might distort the velocity field of ESO~184-G82 via tidal interactions or a major merger.
But interestingly, our detailed study of the spatial and kinematic structure  of the H{\sc i}  in the galaxy shows that the gas is significantly disturbed, with more than 21\% of the gas mass not following the rotation of the main gas disk. 
There are several cases in the literature where galaxies with small companions (with $\sim$10\% of the galaxy mass), or even no detected companions, show peculiar H{\sc i} structure and/or disturbed H{\sc i} kinematics; the disturbed velocity field in such systems is believed to arise from minor mergers 
\citep[for e.g. see][]{Sancisi08}.
This is a likely scenario for ESO~184-G82. The encounter that led to the disturbed H{\sc i} distribution must have taken place relatively recently, given that its tidal effects have not been damped out by the rotation of the galaxy and 
 the disturbed gas has remained rotationally mis-aligned with the main disk. 
This puts an upper limit of a few hundred Myrs, equal to the rotation period of the galaxy disk, on the time of occurrence of the encounter. 
Conversely, the presence of kinematically disturbed gas regions throughout the H{\sc i} disk (see Figure~\ref{fig:chans}) suggest that the encounter occurred sufficiently early on for its tidal effects to disrupt the entire H{\sc i} disk. Indeed, this disruption is likely to have given rise to the W-R region. 
This hypothesis is corroborated by the piling up of high H{\sc i} column density, clumpy gas on the west side of the galaxy, which suggests shock compression of the gas, forming a high density region in which the W-R region and GRB are located, and possibly giving rise to extreme star formation. The disturbed H{\sc i} kinematics and spatial structure of ESO~184-G82 hence suggests that a minor merger in the recent past is the likely cause of the uncommon properties of the W-R region in the galaxy, which, in turn, may have led to the formation of the gamma ray burst. 
Finally, we obtain N(H{\sc i}) $>$ 5 $\times {\rm 10^{20}}$~cm$^{-2}$ \citep[higher than the damped Lyman-$\alpha$ 
threshold of 2 $\times$ 10$^{\rm 20}$ cm$^{\rm -2}$;][]{Wolfe05}, towards the GRB location. 
This is consistent with the high N(H{\sc i}) values typically obtained in absorption studies of GRB host galaxies 
at z $\gtrsim$ 2 \citep[e.g.][]{Fynbo09}. 
Here, with the first H{\sc i}~21\,cm emission mapping of a GRB host, 
we show that the close environment of the GRB is coincident with the highest N(H{\sc i}) region of the galaxy (see Fig.~\ref{fig:maps}). 
To confirm whether  high column densities are typical of  GRB environments, systematic  high resolution H{\sc i}~21\,cm emission mapping 
of a sample of GRB host galaxies is required. 

\section*{Acknowledgments} 
We thank Johan Fynbo, Jayaram Chengalur, and especially Palle M\o ller, for
very helpful discussions and comments. We also thank the referee, Ger de Bruyn,
for a detailed and helpful report. 
NK thanks the DST for support via a Swarnajayanti Fellowship. 
We thank the GMRT staff for having made possible the observations
used in this paper. 
The GMRT is run by the National Centre for
Radio Astrophysics of the Tata Institute of Fundamental Research. 
This paper includes archived data obtained through the Australia Telescope Online Archive.
{Some of the data presented in this paper were obtained from the Mikulski Archive for Space Telescopes (MAST). STScI is operated by the Association of Universities for Research in Astronomy, Inc., under NASA contract NAS5-26555. Support for MAST for non-HST data is provided by the NASA Office of Space Science via grant NNX13AC07G and by other grants and contracts. 

\begin{figure}
\begin{center}
\psfig{file=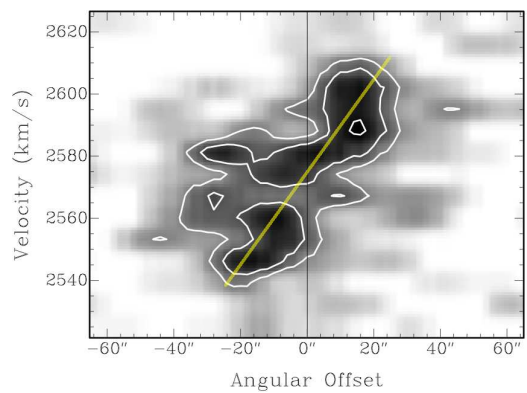,width=2.7truein}
\end{center}
\vskip -3 mm
\caption{
A position-velocity cut through the 25\arcs$\times$ 18\arcs\ spectral cube along the major axis of the galaxy 
(PA: $\sim$130$^\circ$ east of north). The angular offset increases from south-east to north-west, with the galaxy centre at zero angular offset. The outermost contour is at the 3 $\sigma$ level (8.4$ \times {\rm 10^{19}}$~cm$^{\rm -2}$), with each subsequent contour spaced in intervals of $\sqrt{\rm 2}$. Fluxes are measured within a beam centred at each pixel along the p-v cut. The faint yellow line indicates the main rotating gas disk. 
}
\label{fig:pvcut}
\end{figure}

\bsp

\label{lastpage}

\end{document}